# Domain Adaptation of Automated Treatment Planning from Computed Tomography to Magnetic Resonance

Aly Khalifa[1, 2], Jeff Winter[2, 3, 4], Inmaculada Navarro[3, 4], Chris McIntosh[1, 2, 3, 5, 6, 7], Thomas G. Purdie [1, 2, 3, 4]

  [1]  Department of Medical Biophysics, University of Toronto, Toronto, Canada
  [2]  Techna Institute, University Health Network, Toronto, Canada
  [3]  Radiation Medicine Program, Princess Margaret Cancer Center, Toronto, Canada
  [4]  Department of Radiation Oncology, University of Toronto, Toronto, Canada
  [5]  Peter Munk Cardiac Center, University Health Network, Toronto, Canada
  [6]  Joint Department of Medical Imaging, University Health Network, Toronto, Canada
  [7]  Vector Institute, Toronto, Canada

## Abstract

*Objective:* Machine learning (ML) based radiation treatment (RT) planning addresses the iterative and time-consuming nature of conventional inverse planning. Given the rising importance of Magnetic resonance (MR) only treatment planning workflows, we sought to determine if an ML based treatment planning model, trained on computed tomography (CT) imaging, could be applied to MR through domain adaptation.

*Methods:* In this study, MR and CT imaging was collected from 55 prostate cancer patients treated on an MR linear accelerator. ML based plans were generated for each patient on both CT and MR imaging using a commercially available model in RayStation 8B. The dose distributions and acceptance rates of MR and CT based plans were compared using institutional dose-volume evaluation criteria. The dosimetric differences between MR and CT plans were further decomposed into setup, cohort, and imaging domain components.

*Results:* MR plans were highly acceptable, meeting 93.1% of all evaluation criteria compared to 96.3% of CT plans, with dose equivalence for all evaluation criteria except for the bladder wall, penile bulb, small and large bowel, and one rectum wall criteria ($p<0.05$). Changing the input imaging modality (domain component) only accounted for about half of the dosimetric differences observed between MR and CT plans. Anatomical differences between the ML training set and the MR linac cohort (cohort component) were also a significant contributor.

*Significance:* We were able to create highly acceptable MR based treatment plans using a CT-trained ML model for treatment planning, although clinically significant dose deviations from the CT based plans were observed. Future work should focus on combining this framework with atlas selection metrics to create an interpretable quality assurance QA framework for ML based treatment planning.

# Introduction

Machine learning (ML) has the potential to relieve the iterative and time-consuming nature of conventional inverse planning in radiation therapy (RT) by offering increased automation and decision support. When adopted clinically, ML based treatment planning standardizes treatment quality and reduces the overall planning time[1]. While significant research effort has been invested into identifying the optimal ML algorithms[2,3], relatively little attention has been paid to developing a standard quality assurance (QA) framework to determine the robustness of the ML to the diversity of treatment scenarios encountered clinically.

ML based treatment planning models learn the relationship between patients' anatomical features and clinically desired dose distributions by training on a curated dataset of clinical RT plans. Once trained, the model guides treatment planning for a novel by predicting tailored attributes of the dose distribution that would achieve a clinically ideal RT plan. These attributes range from dose-volume constraints for targets and organs at risk (OAR) to the entire voxel-wise dose distribution[2]. These predicted attributes then act as constraints during fluence optimization to create a deliverable plan.

Because ML models are typically trained and validated for a specific clinical scenario, curating training sets of sufficient size and quality for each one is often impractical. Curating medical datasets is challenging due to inadequate data infrastructure, sampling bias, and patient privacy[4]. In the context of radiation therapy, variation in contouring and inherent physician-bias towards achieving certain plan attributes over others additionally complicate curation[5]. To extend the scope of the clinical scenarios that existing trained models can be safely used for in practice, a process called domain adaptation is required. For example, this allows the accommodation of new treatment techniques[6], machines[6–8], patient set-up positions[7], or changes in region of interest (ROI) contouring protocols[9]. It is also desirable for cancers of low incidence where there is not enough data for training a dedicated model, such as adrenal cancers[10].

While more sophisticated methods such as transfer learning can be employed[10,11], often the simplest method of domain adaptation is to reuse the model as-is if the desired treatment plan characteristics in the target domain are similar to the source domain. In this case, the adapted model is typically validated by creating ML treatment plans on a retrospective patient cohort with datasets representative of the target domain and evaluating the model-generated RT plans against the clinically delivered treatment plans. When a ML model makes accurate predictions for previously unseen (i.e., not in the training set) test cases, it is said to generalize well. A model may be unable to generalize when the intrinsic properties of the two domains (i.e., gantry angles, type of modulation, etc.) differ substantially, impacting, for example, high-dose conformality or low-dose spillage[6,7].

In this proof-of-concept study, we investigate domain adaptation in the context of MR based treatment planning. MR-only treatment workflows are gaining popularity to take advantage of the modality's superior soft-tissue contrast and the introduction of integrated MR linear accelerators (MR linac), which offer online adaptation coupled with real-time image guidance during beam delivery. However, there have been relatively few examples where ML based treatment planning has been implemented in the MR guided setting. We therefore investigated if an ML treatment planning pipeline trained on CT imaging could be adapted for use in the MR based treatment planning setting. We utilised a commercially available ML based planning pipeline that uses a wide variety of input features, including ones that are derived from region of interest (ROI) contours and are therefore independent of image intensity values. We therefore hypothesised that the ML based treatment planning pipeline would generate clinically acceptable MR based plans. Additionally, we outline a framework for quantitatively assessing the impact of anatomical variation within the datasets on dosimetric variation.

## Methods

In this study, we use an established and commercially available ML based treatment planning method, trained on CT imaging only, to generate MR based treatment plans. Using retrospectively collected CT and MR imaging from a cohort of prostate cancer patients treated at our institution, we generated CT and MR based RT plans for each patient and compared these plans using our institutional evaluation criteria. We also modeled how anatomical variation within this dataset and the dataset used to train the ML model could contribute to the observed dose differences between CT and MR based plans. Our study design is illustrated in Figure 1 and detailed further below.

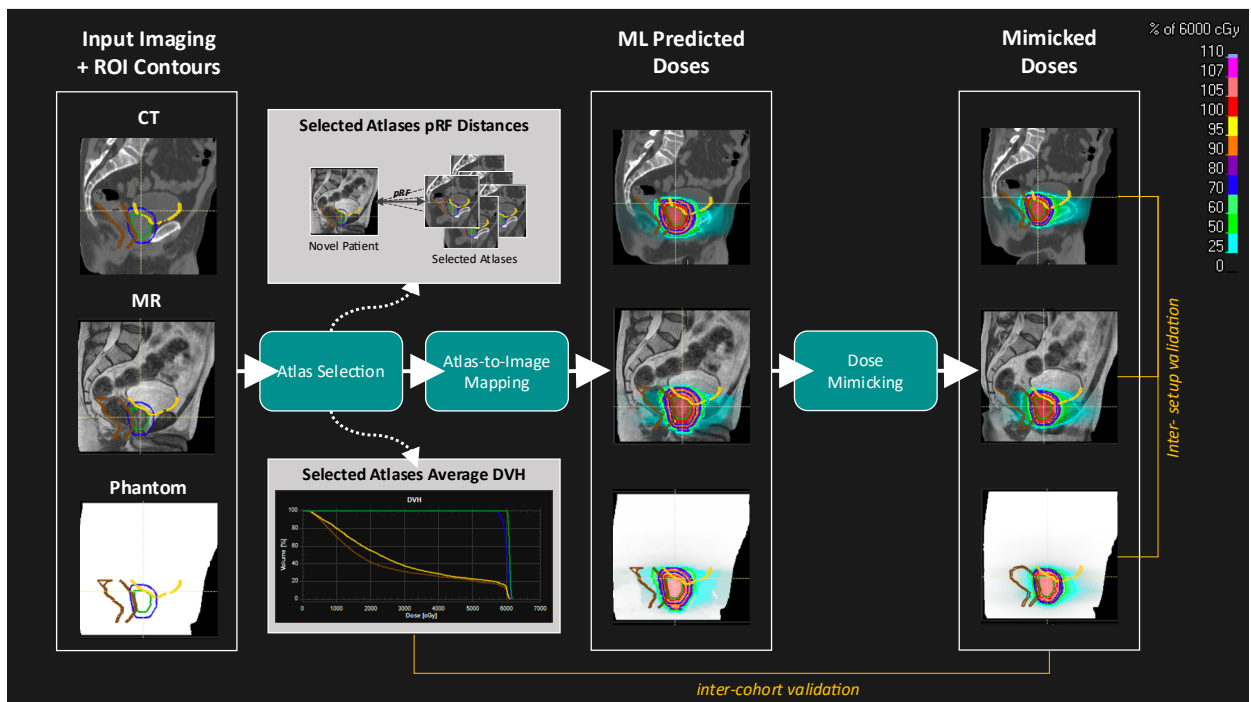

*Figure 1 ML based treatment planning pipeline and study design. The main steps of the ML treatment planning pipeline are atlas selection, atlas-to-image mapping, and dose mimicking. For each patient, a treatment plan was created for each of three imaging modalities (CT, MR, and Phantom). The phantom image was generated by duplicating the ROI contours from the MR image and setting all voxel intensity values to one. The pRF atlas distances are a learned metric used to rank and select the best 5 atlases from the ML model training set. The distances act as a built-in measure of the similarity between an atlas and a novel patient, and are used to create a weighted average dose-volume histogram (DVH) for the selected atlases of each ML treatment plan. The inter-setup validation assesses the dosimetric impact of anatomical differences between the CT and MR/phantom images. Additionally, the inter-cohort comparison assesses the dosimetric impact of atlas-patient mismatch, determined through comparison with the average DVH of selected atlases.*

## Patient Selection and Imaging

We retrospectively collected imaging and contour datasets from 55 prostate cancer patients treated on an MR linac at our institution to determine if of our ML model creates acceptable treatment plans based on MR imaging. These patients were treated with ultra-hypofractionated radiation therapy on the Unity MR linac (Elekta Solutions, Sweden) during one of two prospective clinical trials for MR guided therapy between September 2019 and June 2021. The dataset consists of contoured CT and 3D T2 MR imaging from each patient, acquired on the same day during treatment simulation or as part of the trial setup as per the trial protocols. While similar setup procedures were used before acquiring both images, the bladder preparation protocol for the MR linac simulation was designed to start with a smaller bladder fill to account for bladder filling during treatment on the MR linac. Note that this dataset is independent of the training dataset described below, and we referred to it as the MR linac testing set.

We also generated a third binary 'phantom' image for each patient (example shown in Figure 1), where all patient intensity values of the MR image were set to a uniform intensity value such that anatomical information could only be obtained from ROI contours, which were duplicated from the MR image contours. The phantom image allows us to determine the influence of imaging vs. contour derived features on the predicted dose distribution. As the phantom contours are copied from the MR image, the anatomy of the two images are the same and any differences in the generated treatment plans are independent of anatomical variation.

## Machine learning model

The ML pipeline used has been previously detailed [12,13] and is commercially available in the RayStation 8B treatment planning system (RaySearch Laboratories, Stockholm, Sweden). We used a pre-trained and clinically validated[1,14] model for curative-intent volumetric modulated arc therapy (VMAT) of localized prostate cancer, with a dose prescription of 60 Gy in 20 fractions. The model was trained on a dataset of contoured CT imaging and RT plans from 99 patients. It is FDA, Health Canada, and CE Mark approved, and we therefore relied on this ML pipeline to generate the ground truth CT based RT plans for the comparison. This is in contrast to previous domain adaptation studies that rely on retrospective clinical RT plans for ground truth.

The main steps of the machine learning pipeline are atlas selection, atlas-to-image dose prediction, and dose-mimicking, as illustrated in (Figure 1). The model is atlas based, meaning only the most relevant patients (i.e., atlases) in the training set, as determined by the atlas-selection process, are selected as the basis of dose prediction for a given novel patient. The dose distributions of the selected atlases are mapped onto the novel patient's unique anatomy using a learned dose prediction model specific to each atlas. The individual dose predictions of the top 5 selected atlases are then aggregated to produce a final dose prediction. Finally, dose-mimicking finds the physical radiation machine parameters that best reproduce the predicted dose distribution to create a deliverable plan. These individual steps are detailed further below.

Atlas selection is accomplished by ranking atlases based on a learned distance metric. The metric is learned during a cross-validation procedure, where each atlas is used to predict the dose distribution for every other atlas in the training set. As the clinical doses of the atlases are available, the prediction accuracy can be calculated. The accuracy scores and atlas features form a training set that is used to train a second random forest model, called a pRF, to predict the expected accuracy of each atlas given a novel patient. Each atlas is associated with its own pRF model. During atlas selection for a novel patient, each atlas's pRF predicts an accuracy score for its corresponding atlas, and these scores are used to rank and select the best atlases.

The choice of accuracy metric during pRF training gives rise to two distinct types of atlases: 'spatial' atlases, based on the gamma metric[15] (pass rate at 5%/5 mm) and that are used as the basis of voxel-wise dose predictions; and, 'prior' atlases, based on the mean average difference between dose-volume histograms (DVH) and are used to calculate cumulative dose priors for each ROI. During atlas-to-image dose prediction, the selected prior atlases are used to build a dose prior that enforces the ROI-specific dose-volume constraints observed in the atlases, while the selected spatial atlases guide the spatial distribution of that dose.

For atlas-to-image dose prediction, a contextual (ARF) model, based on the random forest algorithm[16], is used. During the training phase, an ARF for each atlas is trained to predict the voxel-wise dose distribution by learning a mapping between an atlas's features and its clinically delivered dose. These features are either imaging-derived or ROI contour-derived. It is the former imaging-derived features that are sensitive to changes in the input imaging modality, while the latter focus instead on patient geometry. After training, each ARF gives a probabilistic dose distribution for a novel patient. The probabilistic predictions given by the ARFs of the selected spatial atlases are then aggregated using a conditional random field (CRF) optimized over the dose priors per ROI from the selected prior atlases.

Finally, dose-mimicking finds machine delivery parameters that best reproduce the ML predicted dose to ensure a physically realizable and deliverable dose. This is achieved through an optimization algorithm that minimizes the error between the ML predicted voxel-wise dose and the dose distribution to be delivered. Additional optimization penalties are applied to improve target coverage and increase OAR sparing, where possible. These penalties were applied uniformly to all treatment plans in this study.

## Treatment Planning

An ML treatment plan was generated for each patient and image type (CT, MR, and Phantom) in the RayStation treatment planning system (RaySearch Laboratories, Stockholm, Sweden). All plans were for single-arc 360° volumetric modulated arc therapy (VMAT) with a prescription dose of 6000 cGy in 20 fractions to the clinical target volume (CTV). Plan complexity was limited by setting the maximum number of monitor units to 700. Note that this treatment regimen differs from the dose this cohort received clinically to match that of the machine learning model, in this proof-of-concept study.

The CTV, bladder wall, rectum wall, left and right femurs, and penile bulb were contoured as part of routine clinical practice on either the CT or MR reference image for each patient (depending on clinical practice at the time). For this study, an expert radiation oncology fellow retrospectively recontoured all MR and CT images to ensure consistency in contouring practice. The planning target volume (PTV) was created through expansion of the CTV using 7 mm margins posteriorly and 10 mm in all other directions. Additionally, organ wall structures were generated from the bladder and rectum contours through a 3 mm transverse contraction. The wall structures were then trimmed in the inferior and superior directions to within 17 mm of the CTV, as per the PROFIT study[17]. Finally, the large and small bowel contours were generated only within 20 mm of the CTV, which means for some patients these contours were not included.

All dose calculations were performed using the collapsed cone dose engine in RayStation 8B. To permit dose calculation on the MR and phantom images, material density overrides for the CTV, left and right femurs, and external ROI were set for the MR and phantom images using the mean mass density values derived from the corresponding CT image.

Dose mimicking occasionally produces plans with suboptimal target coverage. Clinically, this is accounted for by globally scaling the entire dose distribution to improve target coverage. To replicate this practice in our study, we globally scaled the dose distributions of CT based plans by the minimum amount necessary to meet clinical PTV coverage criteria (PTV D99 > 5700 cGy). If the scaling procedure introduced hotspots (defined as D1cc > 5300 cGy), those plans were only scaled by the amount necessary to maximize PTV coverage without breaking hotspot constraints. To ensure a fair dosimetric comparison between each patient's treatment plans, the MR and phantom based plans were scaled to obtain equivalent PTV coverage to the corresponding CT based plan, regardless of hotspot creation.

## RT Plan Evaluation

Treatment plan quality was compared using dose-volume metrics as per our institutional evaluation criteria (Table 1). The criteria specify dose limits for the CTV, PTV, bladder wall, rectum wall, left and right femurs, penile bulb, and small and large bowel. As the model has been validated for clinical use on CT imaging, the generated CT based plan serves as a control for comparison with the MR and phantom based plans.

Dose equivalence between the CT based plans and the other two plan types was evaluated using paired two-one-sided t-tests (TOST, $\alpha = 0.05$) for each criterion. This tested the null hypothesis that the mean dose difference between groups fell outside of a specified dose threshold, chosen here to be 5% of each criterion's limit. Additionally, statistically significant differences between treatment plan types in the acceptance rates of each criterion were evaluated using Exact McNemar's tests ($\alpha = 0.05$). This tested the null hypothesis that MR and CT based plans have equal probabilities of disagreement with each other (i.e. the probability CT based plan passes a criterion that the corresponding MR based plan does not, and vice versa). False discovery rate correction (FDR < 0.05) for multiple comparisons was applied independently in both cases.

# Evaluation of Dosimetric Variation

We assumed that any observed differences in the dose volume metrics between CT and MR based plans were attributable to one of the following: a) inter-setup variation due to anatomical changes between CT simulation and the first fraction of therapy that could justify differing dosimetric trade-offs, such as organ filling; b) inter-cohort variation due to underrepresentation of MR linac patients' anatomical features in the training set of atlases; or finally, c) inter-domain variation due to altered ML performance as a result of changing the input image modality to one that the model had not been trained on.

As described further below, we quantitively estimated the relative contributions of inter-setup and inter-cohort error using regression analyses and assume that any remaining differences arise from inter-domain error. To facilitate the regression analyses, we quantified anatomical variation in our datasets by measuring the amount of OAR overlap with the PTV as a heuristic anatomical descriptor and surrogate measure of the achievable OAR dose. Specifically, we focused on the PTV-bladder and PTV-rectum overlap percentage as the bladder and rectum are key determinants of clinical acceptability that exist for every plan (unlike bowel). This is in line with previous investigations into the use of PTV-OAR overlap for patient-specific QA to guide and standardize the prostate cancer inverse planning process[18–20]. We assumed that cases with similar PTV-OAR overlap would have similar have the same ability to achieve clinically acceptable treatment plans. The relationship between PTV-OAR overlap and dose-volume metrics was validated using the Pearson correlation coefficient.

*a) Inter-setup validation*: We compared the distributions of PTV-OAR overlap between CT and MR/phantom images to measure the potential for inter-setup error. To quantify the amount of inter-setup error due to anatomical differences between the CT and MR images, we calculated the difference in PTV-OAR overlap, $\Delta V_{setup}$, between the CT image and the other images, and the differences in dose volume metrics, $\Delta D_{setup}$, between their respective treatment plans for each patient. The relationship between these differences was modeled using linear regression:

$$\Delta D_{setup} = m_{setup} * \Delta V_{setup} + b$$

We then used this model to estimate the amount of inter-setup error, $E_{setup}$, expected at the median amount of PTV-OAR overlap difference, excluding the bias term as it would capture the other sources of error:

$$E_{setup} = m_{setup} * median[\Delta V_{setup}] \quad \quad (Equation\ 1)$$

*b) Inter-cohort validation:* We compared the distributions of PTV-OAR overlap between the MR linac cohort and the training set of atlases to measure the potential for inter-cohort error. The distribution of PTV-OAR overlap across both the entire training set of atlases as well as the subset of atlases that were selected for dose prediction was also compared to determine if the pool of atlases the ML could choose from was limited. The relationship between differences in PTV-OAR overlap between each MR linac patient and their selected atlases, $\Delta V_{atlases}$, and the corresponding differences in dose-volume metrics, $\Delta D_{atlases}$, of each was modeled using linear regression. As 5 atlases are selected for each treatment plan, we took a weighted average of the PTV-OAR overlap and dose-volume metrics using the pRF similarity metrics as the weights.

This provides heavier weighting to atlases that the ML deemed more relevant to a patient. For each plan type, we model this relationship as:

$$\Delta D_{atlases} = m_{atlases} * \Delta V_{atlases} + b$$

For each plan type, we then estimate the expected dose change, $E_{atlases}$, at the median amount of atlas-patient PTV-OAR overlap difference as:

$$E_{atlases} = m_{atlases} * median[\Delta V_{atlases}] \text{ (Equation 2a)}$$

Finally, to estimate the inter-cohort error, $E_{cohort}$, we subtract the expected CT dose difference, $\mathbb{E}[\Delta D_{atlases}]_{CT}$, from that of the other modalities.

$$E_{cohort} = E_{atlases,CT} - E_{atlases,I} \quad \text{(Equation 2b)}$$

Where I is the MR or phantom image.

*c) Inter-domain validation*: finally, we assume that any remaining unexplained dosimetric difference is from inter-domain error. To quantify the inter-domain error, $E_{domain}$, we subtracted the expected inter-setup and inter-cohort errors from the observed median difference in dose-volume metrics, $\Delta \mathcal{D}_{observed}$.

$$E_{domain} = \Delta \mathcal{D}_{observed} - E_{cohort} - E_{setup} \quad \text{(Equation 3)}$$

## Atlas Selection

To examine the mechanism by which altering the input image type could lead to dosimetric differences, we assessed changes in the choice of atlases and pRF atlas distances, respectively, as per Conroy et. al.[9]. As the MR and phantom images contain duplicated ROI contours, any differences in the atlases selected are attributable to the altered distribution of image-derived features as a result of changing the imaging modality, rather than changes in patient anatomy. Thus, the frequency that an atlas was selected as either a prior or spatial atlas was quantified in each of the three input cases (CT, MR, and Phantom). Based on this selection frequency, the atlases were then grouped using a hierarchal clustering algorithm to identify atlases with similar selection profiles. This identified how the relative importance of highly selected atlases differed across input modalities. Additionally, relative changes in the distribution of pRF atlas distances for each input case were examined, with prior and spatial atlas distances separately normalized to the minimum and maximum pRF values.

# Results

## Treatment planning

ML based treatment plans were successfully generated for all 55 patients and image types, for a total of 165 plans. In total, the small bowel was contoured for 7 patients on both the CT and MR images, while the large bowel was contoured on the CT of 12 patients and the MR images of 10 patients. Modest improvements in target coverage were possible for most CT based plans using our dose scaling procedure. However, the CT based plans of 4 patients could not be scaled without breaking hotspot constraints.

The doses of MR and phantom plans were then scaled to have equal target coverage as the corresponding CT plan to ensure a common point of standardization. Only one MR based plan could not be scaled without breaking delivery constraints. Overall, the median amount of scaling required for CT based plans was +0.9%, corresponding to a median increase of 49.7 cGy to PTV D99. The dose distributions of MR and phantom based plans were then scaled to match the PTV coverage of the corresponding CT based plan, with median increases of 43.4 and 49.7 cGy to PTV D99, respectively. The dose changes for each ROI due this scaling procedure are shown in Figure S2 Change in dose-volume metrics due to the dose scaling procedure implemented. The procedure multiplicatively scaled the dose distribution of CT based plans to meet PTV coverage criteria, unless hotspots were created. The MR and phantom based plans were then scaled to meet the target coverage of the corresponding CT based plan for each patient..

## RT Plan Evaluation

Dose volume metrics were extracted from all treatment plans (Figure 2) and acceptance rates were determined using institutional evaluation criteria (Table 1). Overall, MR and phantom based treatment plans had high acceptability, meeting 93.1% and 93.3% of all evaluation criteria, respectively. However, they met slightly fewer criteria than CT based plans, which met 96.3% of criteria. The largest dose differences observed were for hotspot criteria, which include the penile bulb and small and large bowel. As a result of the failed hotspot criteria for the small bowel, the acceptance rate was decreased by 14%. However, since the small bowel was contoured for only 7 patients, this decrease is attributable to only one patient.

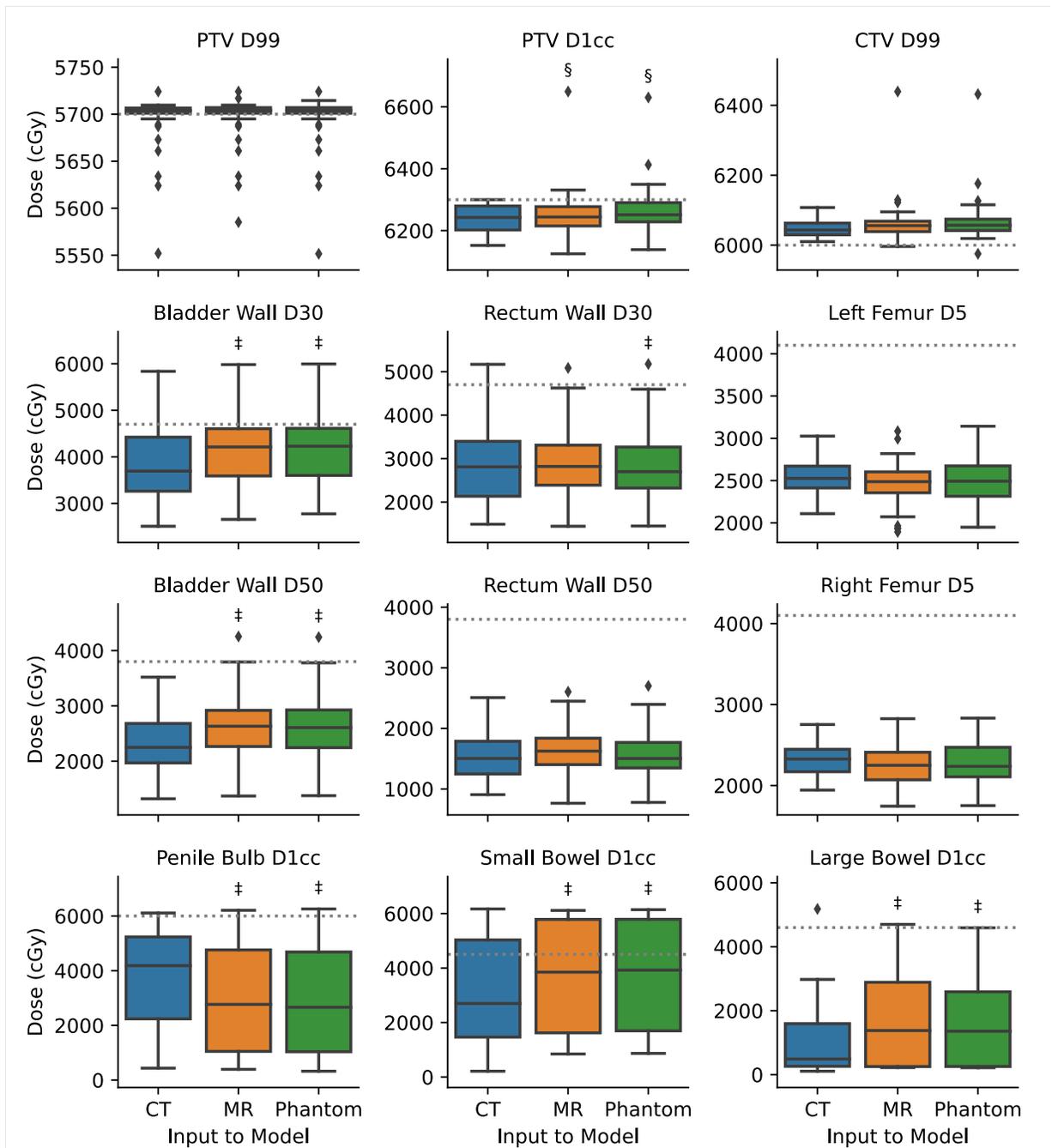

*Figure 2 Dose volume metrics from CT, MR, and phantom based treatment plans. Dotted lines indicate institutional dose evaluation criteria limits. Box plots indicate the median (center line) and interquartile range (IQR). Whiskers denote the minimum and maximum dose values, excluding outliers. Outliers are defined as falling outside 1.5 times the IQR and are shown as diamonds.*

*‡ Denotes plans not found to have equivalent mean dose volume metrics as the CT based plans, identified by TOST tests and FDR correction for multiple comparisons. The equivalence threshold used is 5% of the each criterion's limit.*

*§ Denotes plans with statistically significant differences in acceptance rate compared to the CT based plans, identified by McNemar tests and FDR correction for multiple comparisons.*

The dose differences between treatment plan types for most criteria fell within the specified threshold of equivalence, as determined by TOST tests. However, dose equivalence was not demonstrated for the bladder wall, small and large bowel, penile bulb, and rectum wall D30 criteria. The former two of these ROI were spared less on the MR and phantom based plans, while the latter two had greater sparing. This increased sparing was more pronounced for the penile bulb than the rectum wall. However, none of these dose differences had a statistically significant impact on their acceptance rate, as determined by exact McNemar's tests. Of all criteria tested, PTV D1cc was the only criterion with significant differences in acceptance rates.

|  | Acceptance Rate (%) | | | Median Dose Difference (cGy) $\Delta D_{observed}$ | |
| --- | --- | --- | --- | --- | --- |
| *Evaluation Criteria (cGy)* | CT | MR | Phantom | MR | Phantom |
| CTV D99 > 6000 | 100.0 | 96.4 | 98.2 | 12.6 | 13.7 |
| PTV D99 > 5700 | 85.5 | 85.5 | 85.5 | 0.0 | 1.7 |
| PTV D1cc < 6300 | 100.0 | 83.6§ | 81.8§ | 1.5 | 8.6 |
| Bladder Wall D30 < 4700 | 90.9 | 80.0 | 78.2 | 518.7‡ | 534.5‡ |
| Bladder Wall D50 < 3800 | 100.0 | 98.2 | 98.2 | 381.8‡ | 357.3‡ |
| Rectum Wall D30 < 4700 | 94.5 | 98.2 | 98.2 | 8.2 | -111.8‡ |
| Rectum Wall D50 < 3800 | 100.0 | 100.0 | 100.0 | 120.7 | -1.05 |
| Left Femur D5 < 4100 | 100.0 | 100.0 | 100.0 | -38.6 | -33.1 |
| Right Femur D5 < 4100 | 100.0 | 100.0 | 100.0 | -77.0‡ | -90.2 |
| Penile Bulb D1cc < 6000 | 96.4 | 94.5 | 96.4 | -1416.7‡ | -1521.9‡ |
| Large Bowel D1cc < 4600 | 91.7 | 90.0 | 100.0 | 890.3‡ | 869.6‡ |
| Small Bowel D1cc < 4500 | 71.4 | 57.1 | 57.1 | 1150.6‡ | 1224.8‡ |
| *All Criteria* | **96.3** | **93.1** | **93.3** | - | - |

*Table 1 Acceptance rates of evaluation criteria and differences in dose-volume metrics with respect to the CT based plan. Acceptance rates are calculated as the number of times each criterion was met, divided by the number of times each ROI was contoured.*
‡ *Denotes plans not found to have equivalent mean dose volume metrics compared to the CT based plans, identified by TOST tests FDR correction for multiple comparisons. The equivalence threshold used for TOST is 5% of the clinical limit.*
§ *Denotes plans with statistically significant differences in acceptance rate compared to the CT based plans, identified by exact McNemar tests and FDR correction for multiple comparisons.*

## Evaluation of Anatomical Variation

Anatomical variation between the CT and MR images, as well as population level differences between the MR linac and the training set, all contribute to the observed dosimetric differences between plans, as detailed below. We measured the amount of PTV-OAR overlap for the bladder and rectum wall as a heuristic descriptor for a patient's anatomy and the achievable OAR dose. The amount of PTV-OAR overlap strongly correlated (r > 0.8) with D30 of the rectum and bladder wall, and moderately correlated with D50 (r > 0.6) for the MR linac cohort. For the training set of atlases, the correlation of PTV-OAR overlap with dose-volume metrics was weaker (Figure S1). The relationship between PTV-OAR overlap and dosimetric variation of the final treatment plans is evaluated below in the inter-setup, inter-cohort, and inter-domain validations.

## a) Inter-setup validation

Inter-setup error results from anatomical changes between CT simulation and the first fraction of therapy that could alter the dose trade-offs achievable. We found that both the CT and MR images cover similar ranges of PTV-bladder overlap (9 – 27%) and PTV-rectum overlap (0 – 25%), although the MR images had a higher median PTV-bladder overlap and lower median PTV- rectum overlap compared to the CT images (Figure S1).

Dosimetric differences between CT based and MR based plans were proportional to the anatomical differences between them (Figure 3). For the bladder wall and rectum wall, D30 moderately correlated ($r > 0.7$) and D50 weakly correlated ($r > 0.5$) with the differences in overlap between images. The median amounts of PTV- bladder overlap and PTV-rectum overlap difference were, respectively, 0.6% and -0.7%. The expected dose difference due to anatomical differences, as predicted by the linear regression model, are shown in Table 2. Overall, the expected median dose differences are small – less than about 75 cGy. Additionally, several patients exhibited large differences in the PTV-OAR overlap that resulted in large dose differences between treatment plans. For example, the largest difference in PTV-rectum overlap observed was 16%, which corresponded to as much as a 10 Gy difference in D30 between treatment plans.

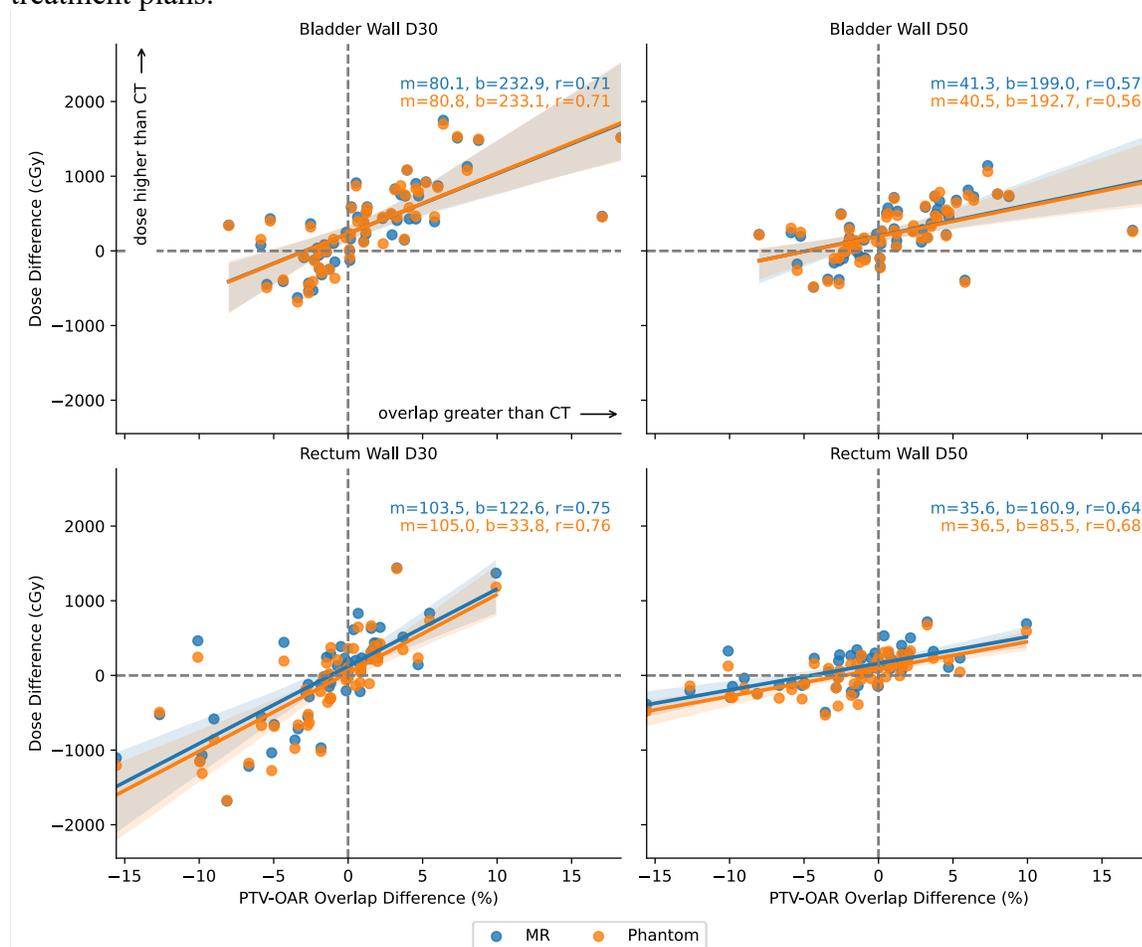

Figure 3 The relationship between differences in dose-volume metrics and in PTV overlap percentage. The slope (m) and intercept (b) of the regression lines, as well as the Pearson correlation coefficient (r) are indicated.

|  | Observed Median PTV Overlap Difference (%) $median[\Delta V_{setup}]$ | | Inter-setup Error (cGy) $E_{setup}$ | |
|---|---|---|---|---|
|  | MR | Phantom | MR | Phantom |
| Bladder Wall D30 | 0.63 | 0.63 | 50.8 | 51.2 |
| Bladder Wall D50 | 0.63 | 0.63 | 26.2 | 25.7 |
| Rectum Wall D30 | -0.73 | -0.73 | -75.1 | -76.2 |
| Rectum Wall D50 | -0.73 | -0.73 | -25.8 | -26.5 |

*Table 2 The median amount of PTV-OAR overlap difference between the CT image and other image types. The expected dose due to inter-setup error was calculated by multiplying the median overlap difference by the slope coefficient from the regression analysis (Equation 1).*

### b) Inter-cohort validation

Inter-cohort results from altered ML performance due to underrepresentation of MR linac patients' anatomical features in the training set of atlases. We compared the distributions of PTV-OAR overlap for all atlases available for selection from the training set (n = 99) and the subset of atlases selected during treatment planning (n=49). The available training atlases represent only a subset of the PTV-OAR overlap range covered by the MR and CT images, excluding more extreme values of overlap (>21 % for the bladder wall, and <6% and >21% for the rectum wall). The selected atlases cover a range of PTV-OAR overlap comparable to the available atlases, but only a subset of the range covered by the MR linac cohort (Figure S1).

We found that the amount that a ML plan deviated from the dose of their selected atlases was proportional to the PTV-OAR overlap discrepancies between them (Figure 4). This relationship strongly correlated (r > 0.8) for the rectum wall and moderately correlated for the bladder wall (r > 0.7). The median amount of PTV-OAR overlap difference between ML plans and their selected atlases for the MR and phantom plans was nearly double that of the CT plans, and this corresponded to an increased bladder and decreased rectum dose on the MR and phantom plans, compared to the CT plans (Table 3).

|  | Observed Median PTV Overlap Difference (%) $median[\Delta V_{atlases}]$ | | | Expected Dose Difference from Atlases (cGy) $E_{atlases}, I$ | | | Inter-cohort error (cGy) $E_{cohort}$ | |
|---|---|---|---|---|---|---|---|---|
|  | CT | MR | Phantom | CT | MR | Phantom | MR | Phantom |
| Bladder Wall D30 | 1.40 | 2.68 | 2.76 | 165.3 | 345.0 | 368.5 | 179.7 | 203.2 |
| Bladder Wall D50 | 1.40 | 2.68 | 2.76 | 99.0 | 232.1 | 243.3 | 133.1 | 144.3 |
| Rectum Wall D30 | -2.61 | -5.61 | -4.74 | -300.6 | -486.1 | -393.9 | -185.5 | -93.3 |
| Rectum Wall D50 | -2.61 | -5.61 | -4.74 | -112.5 | -220.1 | -169.1 | -107.6 | -56.6 |

*Table 3 Median differences in PTV overlap between ML plans and their selected atlases. The expected dose increase due to atlas-patient mismatch was estimated by multiplying the median amount of PTV overlap by the slope term from the regression analysis (Equation 2a). The expected dose difference between CT and MR/phantom plans due to inter-cohort error was calculated by subtracting the expected doses of the MR/phantom plans from those of the CT based plans (Equation 2b).*

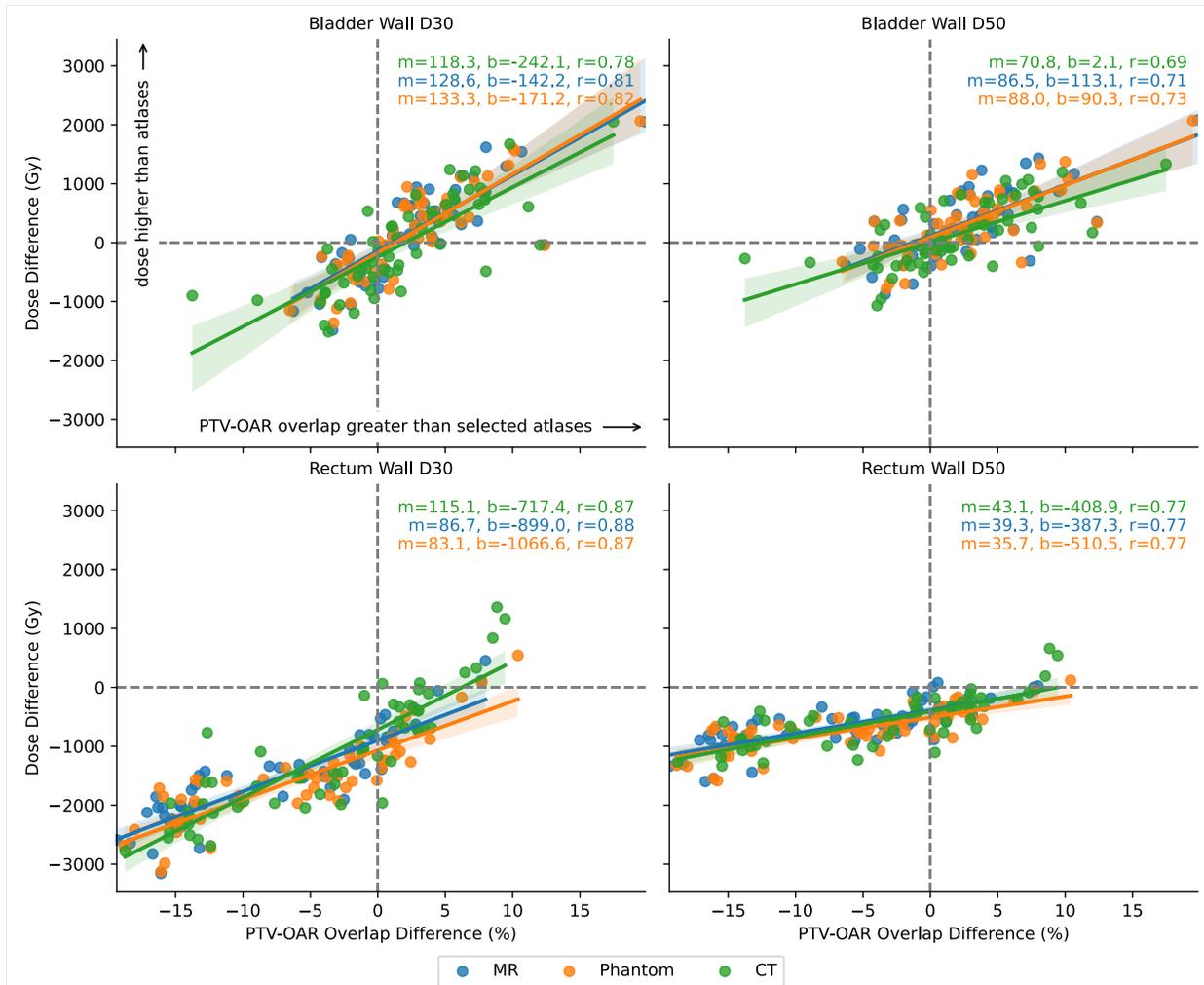

*Figure 4 The relationship between differences in PTV overlap and in dose-volume metrics with respect to ML plans and their selected atlases. The slope (m) and intercept (b) of the regression lines, as well as the Pearson correlation coefficient (r) are indicated.*

### c) *Inter-domain validation*

Inter-domain variation results from altered ML performance as a result of altering the input feature distribution to one that the model had not been trained on. We assumed that any remaining dose unexplained by inter-setup or inter-cohort variation is due to inter-domain variation (Equation 3). Table 4 summarizes the contribution of each source of variation. For both the bladder and rectum wall, inter-domain variation causes the dose of MR and phantom plans to increase. For the bladder wall, inter-domain variation is responsible for more than half of the observed dose difference between MR and CT based plans. For the rectum wall, our model predicts that the total inter-domain variation is greater than the observed dose. This indicates that the inter-domain variation is counteracted by the inter-setup and inter-cohort variation, both of which predicted increased sparing of the rectum wall (Table 4).

| Metric | Observed Difference in Median Doses from CT Based Plan (cGy) $\Delta \mathcal{D}_{observed}$ | | Inter-setup Variation (cGy) $E_{setup}$ | | Inter-cohort Variation (cGy) $E_{cohort}$ | | Inter-domain Variation (cGy) $E_{domain}$ | |
|---|---|---|---|---|---|---|---|---|
| | MR | Phantom | MR | Phantom | MR | Phantom | MR | Phantom |
| Bladder Wall D30 | 518.7 | 534.5 | 50.8 | 51.2 | 179.7 | 203.2 | 288.2 | 280.1 |
| Bladder Wall D50 | 381.8 | 357.3 | 26.2 | 25.7 | 133.1 | 144.3 | 222.5 | 187.3 |
| Rectum Wall D30 | 8.2 | -111.8 | -75.1 | -76.2 | -185.5 | -93.3 | 268.8 | 57.7 |
| Rectum Wall D50 | 120.7 | -1.05 | -25.8 | -26.5 | -107.6 | -56.6 | 254.1 | 82.05 |

*Table 4 The individual contributions of inter-setup, inter-cohort, and inter-domain variations to the observed dose difference between MR/phantom plans and CT based plans. Inter-domain variation was calculated using Equation 3.*

## Atlas Selection

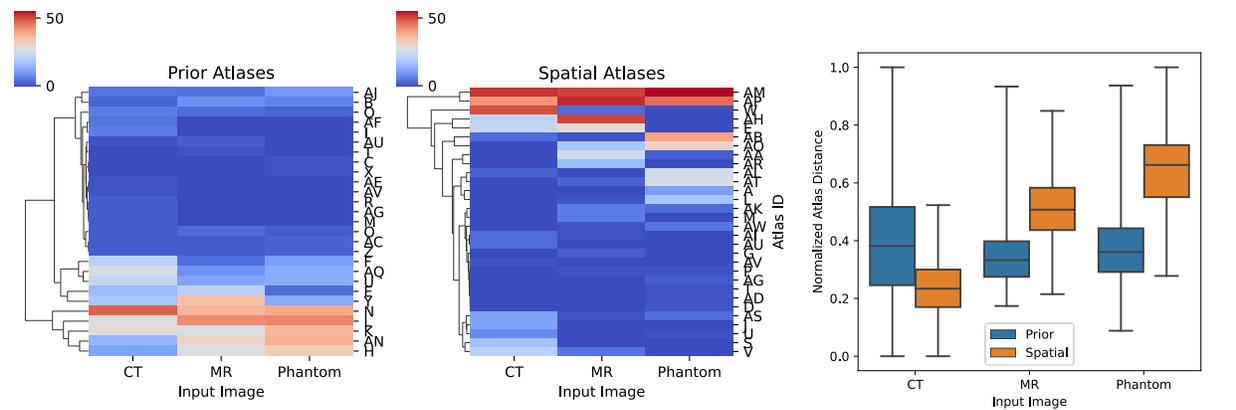

*Figure 5 Cluster map showing the frequency that an atlas was selected as either a prior (left) or spatial (middle) atlas for each of the image types. Boxplot shows the distribution of pRF atlas distances, normalized to the minimum and maximum pRF distances independently for both atlas types (right).*

Hierarchal clustering of the atlas selection frequency data reveals that the atlas selection process behaved differently in each of the input cases, both in terms of the specific atlases selected and the pRF atlas distances (Figure 5). Consequently, a different subset of all available atlases was selected in each input case. While some atlases that are frequently selected in CT plans are also frequently selected in the MR and phantom plans, there are several examples where these atlases are less frequently selected. Despite having duplicate ROI contours, there are also cases where the atlases selected in the MR plans differ from the phantom plans. Additionally, the PRF values in both the MR and phantom plans were elevated in comparison to those of the CT plans.

# Discussion

With the rising importance of MR as imaging modality in radiation oncology workflows, our study investigated if a treatment planning ML model trained on CT imaging could be repurposed for MR based planning. We generated treatment plans on matched CT and MR images from patients treated on our institution's MR linac, and quantitively compared RT plan quality by evaluating changes in dose-volume metrics and acceptance rates, according to our institutional evaluation criteria. We additionally implemented a regression analysis to differentiate the observed dosimetric differences into three sources: 1) inter-setup variation, caused by anatomical differences between CT simulation and the first fraction of therapy; 2) inter-cohort variation, caused by underrepresentation of the MR linac cohort in the ML model training set; and, 3) inter-domain variation, caused by incorrect model predictions due to an underlying change in the input feature distribution.

Our repurposed ML model was able to generate highly acceptable MR based treatment plans, meeting more than 93% of evaluated clinical criteria (Table 1). However, several important dosimetric differences were observed between the dose-distributions of the MR and CT based RT plans. The bladder wall was most heavily impacted OAR, with MR based RT plans having a higher average dose to the bladder wall than the CT based plans, resulting in 15% more treatment plans exceeding the clinical limit. While no similar dose differences were observed for the rectum, this seemed to be the result of competing sources of dosimetric variation. Our analysis predicted that both inter-setup and inter-cohort validations would lead to increased rectum sparing, while the inter-domain validation predicted less sparing - canceling each other out.

We implemented bladder and rectum overlap percentages (PTV-OAR overlap) as heuristic anatomical descriptors and surrogate measures of the achievable OAR dose, which have been previously studied as a QA tool in conventional inverse planning[18–20]. As expected, we found that the amount of PTV-OAR overlap correlated with dose-volume metrics (Figure S1). Although ML based planning is similarly designed to predict clinically achievable and acceptable doses, the predictions are difficult to interpret or explain, making it difficult to justify them in new domains. Our intention was therefore not to replace the ML predicted dose, but rather develop a secondary, more interpretable, heuristic that is independent of model performance. By measuring differences in PTV-OAR overlap and dose-volume metrics across RT plans, we were able to estimate the relative contributions of inter-setup, inter-cohort, and inter-domain variation.

We found that inter-domain variation was the primary source of dosimetric variation, accounting for more than half of the dosimetric differences observed between CT and MR RT plans (Table 4). This result was expected as our ML model partially depends on image-derived features to make predictions. As only CT images were seen during training, the learned mapping between image features and dose are specific to CT. Using MR images as input alters the distribution of these features and the model is consequently unable to correctly map to dose. Despite this, we were able to generate acceptable RT plans for a large portion of the patients in our cohort, suggesting that contour-derived features are sufficient for accurate prediction of the dose. While it was previously shown that inclusion of contour-derived features can improve model performance[14], our results support that it may be possible to exclude imaging-derived features all together.

Inter-cohort variation was the second largest contributor to dosimetric differences, which is a result of population-level anatomical differences between the training cohort and MR linac. We found that the more extreme values of PTV-OAR overlap of the MR linac cohort were not represented in the training set (Figure S1), consequently limiting the availability of closely matching atlases for the MR linac cohort. We found that the amount of PTV-OAR overlap difference strongly correlates with the dose differences between the atlases and the ML predicted dose, even in regions of PTV-OAR overlap that are not represented by the training set (Figure 4). Our regression analysis predicted that patients would have higher bladder and lower rectum median doses on their MR plans. This discrepancy between cohorts could be a result of differing patient inclusion criteria or patient set-up procedures.

The linear relationship between PTV-OAR overlap difference and dose difference between patients and atlases is likely because the dose prior for the overlap region models the voxel-wise dose independently of the size of the overlap region. Thus, any change in the size of the overlap region with respect to the selected atlases results in a proportional change in dose-volume metrics. Fortunately, dose differences align with what would be expected clinically – that less overlap with the PTV indicates a more favourable case and vice versa. However, in the absence of a clinical treatment plan as a ground truth for comparison with the MR based plans, it is unclear if the resulting dose is providing maximally achievable OAR sparing.

Finally, we found that inter-setup variation contributed only a modest amount to the dosimetric difference. Inter-setup variation results from anatomical discrepancies between the CT and MR images, likely due to discrepancies in patient preparation practices between the CT simulation and MR linac simulation imaging in this study. The MR linac treatment sessions are longer due to the adapted RT steps, and a smaller bladder is required for the initial image to ensure patient comfort with the expectation that bladder filling will occur over the course of treatment. Another example could be difference in bladder or bowel preparation between RT simulation and the first treatment fraction, which is a potential issue for any radiation treatment process. While ideally these should be equivalent to minimize set-up error, it is natural for minor differences in workflows to occur. Additionally, the increased bladder volume on CT may also explain why MR plans had increased bowel doses (Figure 2), as the bladder tends to move the bowel regions further from the region of high dose near the target as it fills (Figure S3) and D1cc criteria are sensitive to even small anatomical differences.

Disruption of the atlas-selection process also appears to be a source of dosimetric variation in this study. We identified clusters of atlases that were highly selected for the CT based plans but not for the MR or phantom based plans, and vice versa (Figure 5). Additionally, the pRF atlas distances are elevated in the MR and phantom cases, which may be a result of both the anatomical differences between cohorts and the change in input feature distribution (i.e., MR imaging instead of CT). As the pRF distances are also a learned metric, they are also sensitive to changes in the input feature distribution. For example, even though MR and phantom images were anatomical duplicates, the atlases selected in each case differed. Thus, elevation in atlas distances may also signify inter-domain effects, in line with previous studies[1,9] demonstrating that these metrics are sensitive in changes in the input and may serve as a basis for predicting the applicability of the ML model for a given patient.

Our main contribution is to propose a QA framework for domain adaptation of pre-trained ML based treatment planning solutions. Differentiating between the sources of dosimetric variation and quantifying them provides actionable information for clinical teams hoping to reuse existing models. For example, specifying a threshold for the acceptable amount of inter-domain variation allows teams to determine whether models can be safely adapted to the target domain or if retraining the model on a dataset specific to that domain is necessary. Additionally, the amount of inter-cohort variation measures if a model's training set was representative of cases in the target domain, while large inter-setup variation could prompt a team to revise its patient setup practices.

Modifying the dose-mimicking strategy may also allow teams to compensate for observed inter-domain variation. We applied dose-mimicking uniformly across all three plan types in our study to ensure a fair comparison, but clinically these can be fine-tuned to prioritise specific dose objectives. For example, the mimicking strategy could be modified to focus on increasing bladder wall sparing to counteract the elevated doses observed. This provides greater flexibility for fine-tuning a model to a specific application domain without the need for model retraining.

Although we knew beforehand that the main source of variation was the change in input modality, we had no prior knowledge that there would be anatomical differences between cohorts. These differences resulted in additional error when adapting the model to the MR guided domain. This highlights the importance of designing QA systems that can detect unexpected changes in the inputs as testing every possible source of variation is impractical.

Atlas based methods are uniquely advantageous in that they provide two main avenues for explaining dose predictions: the intrinsic pRF atlas distances, which serve as a built-in measure of prediction confidence; and the ability to directly compare a predicted dose distribution to its selected atlases. These results demonstrate that significant differences between a novel patient's anatomy and their selected atlases' indicates that that the model may not be suitable for that patient. Thus, the utility of atlas distance metrics could be enhanced by complementing them with anatomical descriptors, such as PTV-OAR overlap, to facilitate atlas-patient comparison and provide an additional layer of interpretability to guide clinical decision making.

A limitation of our study is that the overlap percentage is only applicable to OAR that overlap with the PTV, potentially missing the dose relationships of other critical organs. Even for the rectum, we observe several cases where there is no overlap due to the use of SpaceOAR Hydrogels (Boston Scientific, United States) implanted to move the rectum away from the PTV (Figure S1). Additionally, PTV-OAR overlap does not fully capture the geometry of the overlap region, such as to what degree an OAR wraps around the target. We chose this metric for its conceptual and implementational simplicity, but future work should focus on more sophisticated heuristics that better characterise the OAR's position and geometry with respect to the target, such as the overlap volume histogram[21].

Another limitation is our use of whole-organ density overrides on the MR images to permit dose calculation. This method was, again, selected for its simplicity but may not achieve the same level of accuracy as other synthetic CT techniques[22]. Density overrides may overlook salient aspects of the image that would affect the dose distribution, such as the presence of air pockets. Accounting for air pockets may be especially important in the MR-guided domain, where the magnetic field impacts the dose at air-tissue interfaces via the electron-return effect[23]. While image-derived features could potentially account for their presence, this would require the inclusion of MR imaging in the ML model training dataset.

Future work in this area should focus on developing and integrating anatomical descriptors to model QA procedures, including a larger set of OAR and treatment sites. As our plan evaluation methods do not account for other indicators of plan quality such as conformality, complexity, or physician preference, these results should be validated by blinded review from physicians and physicists. This will ensure the system aligns with clinicians' judgment, instead of only relying on technical measurements. Finally, the framework provided here should be integrated with atlas selection metrics into a more holistic framework for patient-specific QA as part of daily clinical practice.

# Conclusion

In this study, we demonstrated that our ML based treatment planning method produces clinically acceptable MR based RT plans, despite being trained on CT imaging only. We provide a novel framework for quantifying anatomical variation within datasets and their dosimetric impact when adapting a pre-trained model to a new application domain. While inter-domain variation contributed the most to dosimetric differences between domains, inter-cohort variation was also a large contributor, highlighting the need for increased attention when curating training sets for new application domains. Future work should focus on combining this framework with atlas selection metrics to create an interpretable QA framework for ML based treatment planning.

# Ethical statement

MR linac patient data was acquired retrospectively from two clinical trials approved by our institutional Research Ethics Board (REB). REB approval for this study was waived by our institutional Quality Improvement Review Committee as part of an ongoing quality improvement initiative for artificial intelligence-based treatment planning.

# Supplementary Material

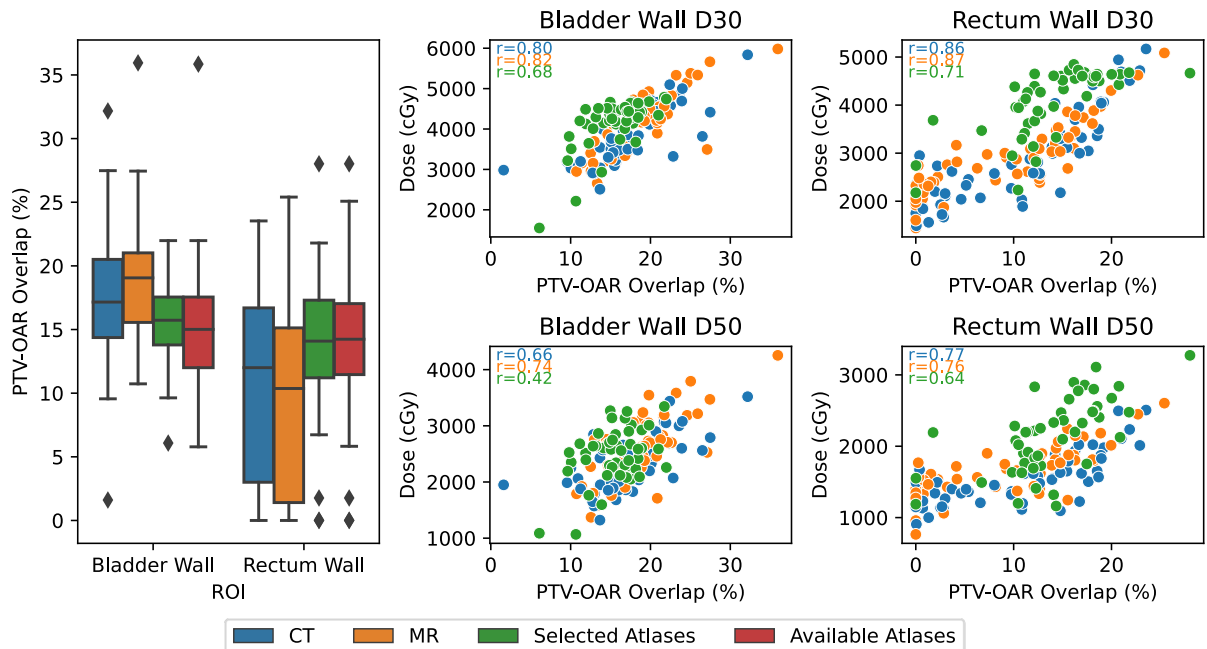

*Figure S1 Distribution of bladder and rectum wall overlap percentage (PTV-OAR overlap) for the CT and MR images of the MR linac cohort, and for the available and selected atlases (left). Pearson correlations (r) of the PTV-OAR overlap with dose-volume metrics for the bladder wall and rectum wall (right). PTV-OAR overlap values for phantom images are not shown as they are equal to those of the MR image.*

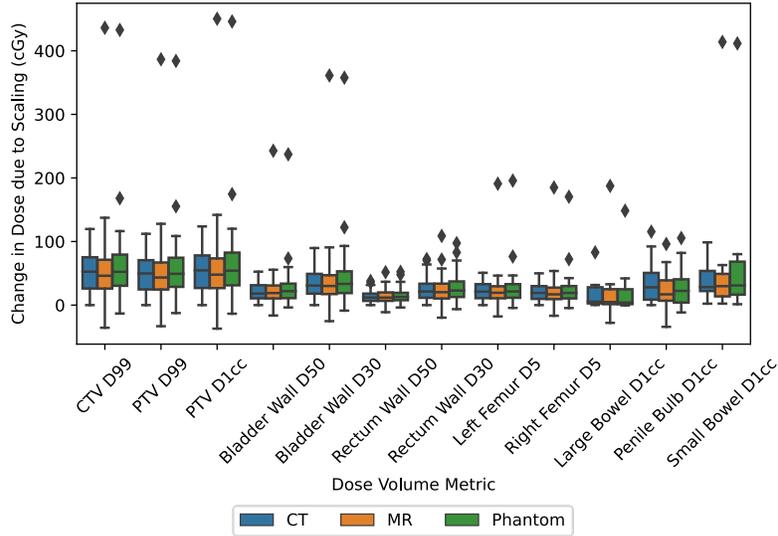

*Figure S2 Change in dose-volume metrics due to the dose scaling procedure implemented. The procedure multiplicatively scaled the dose distribution of CT based plans to meet PTV coverage criteria, unless hotspots were created. The MR and phantom based plans were then scaled to meet the target coverage of the corresponding CT based plan for each patient.*

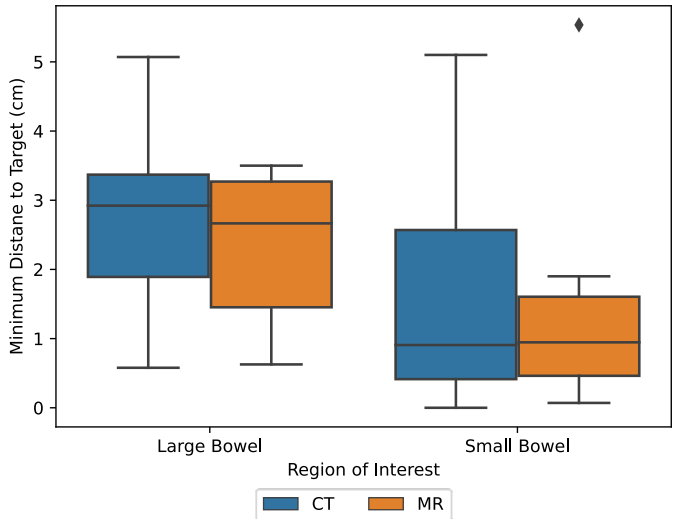

*Figure S3 The minimum distances of the small and large bowels to the clinical target volume (CTV) surface from patients who had bowel contours. The minimum distance was obtained from the signed distance transform, which calculates the minimum distance to the CTV surface for each voxel of the bowel ROI.*